\documentclass[11pt]{article}
\usepackage{moriond,epsfig}

\def\be{\begin{equation}}
\def\ee{\end{equation}}
\def\bea{\begin{eqnarray}}
\def\eea{\end{eqnarray}}

\newcommand\joref[5]{#1, #5, {#2, }{#3, } #4}

\def\hMpc{\mbox{h$^{-1}\,$Mpc}}
\def\aanda{A\&A}                      
\def\cqg{Class. Quant. Grav.}         
\def\mnras{MNRAS}     

\begin{document}
\vspace*{4cm}
\title{
CLUSTERS OF GALAXIES AS STANDARD CANDLES FOR GLOBAL OBSERVATIONAL COSMOLOGY
}

\author{ B.F. ROUKEMA }

\address{DARC/LUTH, Observatoire de Paris--Meudon, 
5, place Jules Janssen, F-92195 Meudon Cedex, France}

\maketitle\abstracts{
As the largest gravitationally collapsed objects, and as
objects with a relatively low space density, clusters of galaxies
offer one of the best sets of standard candles for trying to measure
basic cosmological parameters such as the injectivity diameter
$2r_{\mbox{inj}}$ (the shortest distance between two topologically lensed
images of any object, e.g. cluster) and the out-diameter $2r_{+}$ (the
maximum `size' of the Universe). Present constraints indicate that
either of these may be smaller or larger than the horizon diameter.
}


\section{Global observational cosmology versus local observational
cosmology}

As was noticed by Karl Schwarzschild\cite{Schw00}, the three-dimensional 
space that we live in may or may not be curved, and
it may or may not be multiply connected. Two-dimensional examples of flat 
and curved spaces include the infinite flat plane and (the surface of) 
a sphere, respectively; 
examples of simply connected and multiply connected spaces 
include the infinite flat plane and a cylinder, respectively.

Curvature is a local property; connectedness (topology) is a global property.

The advent of general relativity and the
Friedmann-Lema\^{\i}tre-Robertson-Walker (FLRW) model led to little change in
empirical constraints on these properties for most of the twentieth
century. For a recent general introduction to the subject, see
Luminet \& Roukema\cite{LR99} or Roukema\cite{Rouk00b}. 

Different definitions of the size of an FLRW
universe are possible and are presented in \S5.1 of ref.\cite{LR99}, 
in particular
the {\em injectivity diameter}
$2r_{\mbox{inj}}$ (the shortest distance between two topologically lensed
images of any object, e.g. cluster) and the {\em out-diameter} $2r_{+}$ (the
maximum `size' of the Universe). For example, the injectivity diameter
of a cylinder universe is its circumference, but its out-diameter is infinite.

Constraining $2r_{\mbox{inj}}$ and $2r_{+}$ is a fundamental goal
of observational cosmology. The basic principle underlying nearly all
methods of constraining these parameters is the existence of 
multiple, topologically lensed images of a single object (e.g. think
of the multiple spatial geodesics on a cylinder joining an arbitrary
point $A$ to the observer at $B$), where the object is either
a physically collapsed extragalactic object such as a cluster of
galaxies or a quasar, or a patch of plasma which emits microwave background
radiation.

For a compact and recent classification of
the numerous different approaches, see ref.\cite{Rouk01}.

\section{Using X-ray clusters to constrain local observational
cosmology}

Several contributors to this meeting have presented work on using
observations of clusters of galaxies, particularly as detected in X-rays,
in order to see to what extent these can or cannot 
constrain local cosmological parameters such as the matter density 
parameter $\Omega_{\mbox{m}}$ and the cosmological constant 
$\Omega_\Lambda$.

\section{Using X-ray clusters to constrain global observational
cosmology}

Gott\cite{Gott80} found that the lack of topologically lensed images 
of the Coma cluster closer to the observer than Coma itself implied
a minimum value of $2r_+ \approx 60${\hMpc}.

Lehoucq, Luminet \& Lachi\`eze-Rey\cite{LLL96} 
used a catalogue of Abell and ACO clusters to
demonstrate a statistical method of searching for pairs of 
topological images in a catalogue with negligible selection effects.

Roukema \& Edge\cite{RE97} 
pointed out that since the formation time of
the most massive X-ray clusters is similar to the age of the Universe,
these objects should be much closer to a ``standard candle'' (once
formed)
useful 
for searching for multiple topological images than other extragalactic
objects, such as quasars.

Using archival data, Roukema et al.\cite{RE97,RB98,RBa99,Rouk00a} 
have investigated a serendipitous hypothesis
for the global topology of the Universe, corresponding to one of
the simplest multiply connected models possible, a ``2-torus'' model,
in which the Coma cluster, and the clusters 
RX~J1347.5-1145 and CL~09104+4109 are three images of a single, physical
cluster. Since the archival data is of insufficient quality to provide
strong constraints, an observational programme is underway 
in order to provide
serious constraints on this working hypothesis.

The arguments of Gott\cite{Gott80} and 
Roukema \& Edge\cite{RE97} remain valid for all
future candidate 3-manifolds (``topologies''): possibly the best
way to test a candidate 3-manifold
detected in cosmic microwave background data from the MAP and/or
Planck satellites will be to 
conduct X-ray searches for clusters of galaxies at the angular positions
and redshifts predicted by the candidate model. The model should then be
quickly confirmed or falsified given sufficient telescope time.



\end{document}